\documentclass[10pt,letterpaper]{article}
\usepackage{opex3}
\usepackage{cite}

\begin{document}

\title{Efficient and mode selective spatial mode multiplexer based on multi-plane light conversion}

\author{Guillaume Labroille,$^1$ Bertrand Denolle,$^1$ Pu Jian,$^{1,*}$ Philippe Genevaux,$^2$ Nicolas Treps,$^{1,3}$ and Jean-Fran\c{c}ois Morizur$^1$}
\address{$^1$CAILabs, 8 rue du 7e d'Artillerie, 35000 Rennes, France \\
$^2$Alcatel-Lucent Bell Labs, Route de Villejust, 91620 Nozay, France \\
$^3$Laboratoire Kastler Brossel, Universit\'e Pierre et Marie Curie, ENS, CNRS, \\
4 place Jussieu, 75005 Paris, France}

\email{contact@cailabs.com} 



\begin{abstract}
We designed and built a new type of spatial mode multiplexer, based on Multi-Plane Light Conversion (MPLC), with very low intrinsic loss and high mode selectivity. In this first demonstration we show that a typical 3-mode multiplexer achieves a mode selectivity better than $-23$~dB and a total insertion efficiency of $-4.1$~dB (optical coating improvements could increase efficiency to $-2.4$~dB), across the full C-band. Moreover this multiplexer is able to perform any mode conversion, and we demonstrate its performance for the first 6 eigenmodes of a few-mode fiber: LP$_{01}$, LP$_{11\mathrm{a}}$, LP$_{11\mathrm{b}}$, LP$_{02}$, LP$_{21\mathrm{a}}$ and LP$_{21\mathrm{b}}$.
\end{abstract}


%
%

\bibliographystyle{osajnl}

\section{Context}

Single-mode-fiber-based transmission systems are rapidly approaching their capacity limit~\cite{essiambre2012} and the research community is working on space-division multiplexing over novel fiber types as a solution against the capacity crunch~\cite{richardson2013}. One possible approach for space-division multiplexing is to use multicore fibers, where the data is multiplexed over the uncoupled cores of the fibers~\cite{hayashi2011,takara2012}. In this article, we are interested in another approach, namely mode-division multiplexing, where the spatial modes of a few-mode fiber (FMF) are used as independent data channels~\cite{ryf2012,salsi2012}. It is also possible to exploit high order orbital angular momentum modes~\cite{fontaine2012b,wang2012} in vortex fibers~\cite{bozinovic2013}.

The realization of a full mode multiplexed transmission system with a FMF~\cite{ip2013,sleiffer2012} requires the development of specific components. One of the key elements in mode-division multiplexing is the spatial multiplexer (MUX) and demultiplexer (DEMUX) allowing to combine and separate the channels in and out of a FMF. More precisely, the spatial MUX combines $N$ input signals from single-mode fibers (SMFs) onto $N$ independent data channels in the FMF, and conversely the spatial DEMUX separates these channels after transmission in the FMF. 


Two strategies exist in order to multiplex the input channels in such a way that they can be demultiplexed at the output. On the one hand, one can launch the input channels into a set of orthogonal combinations of the eigenmodes of the FMF. The channels are thus strongly coupled, and demultiplexing the channels therefore requires full modal diversity at the receiver and complex digital signal processing, especially when the fiber exhibits large differential group delay and mode differential loss. Current coherent receivers operating over single-mode fiber have a $2\times 2$ multiple input multiple output filter to manage the two states of polarization of the light~\cite{savory2010}. The depth of each filter is linked to the polarization mode dispersion of the fiber and is around 5 to 10 symbols. By comparison, a 10-mode transmission would require an array of 10 synchronous coherent receivers and a $20\times 20$ multiple input multiple output filter with a depth typically on the order of 100 symbols for a 30~km transmission~\cite{randel2011}. On the other hand, one can use fibers in which spatial modes are weakly- or un-coupled. In this case, if one launches each channel into an individual eigenmode of the fiber, only the degenerate mode groups must be jointly detected, and nearly standard signal processing at the receiver side can be used.


Here we focus on this second technique. For weakly coupled system, each network element along the transmission path needs to have low mode crosstalk. The accumulated crosstalk needs to stay below a given limit, depending on modulation format. This puts a strong requirement on mode coupling for each individual component. Thus one of the challenges is to design a mode MUX/DEMUX with high mode selectivity and low insertion loss. 

Up to now several designs have been used to build mode MUXes~\cite{ryf2012b,fontaine2012,chen2012,leuthold1996,ding2013,li2013,love2012,driscoll2013,luo2014}. A conventional solution is to use binary phase plates to convert single mode LP$_{01}$ from a single mode fiber to a higher order linearly polarized (LP) mode~\cite{ryf2012b}. Each mode is generated with a specific phase plate and all the generated LP modes are superposed together into the FMF fiber with a succession of beamsplitters. The drawback of this design is that the intrinsic loss due to beamsplitting scales up with the number of mode. For instance, a MUX with 4 modes imposes 6~dB intrinsic superposition loss, and a MUX with 8 modes imposes 9~dB loss, on top of all other conversion losses. 

Another technique to build a mode MUX is to use a photonic lantern~\cite{fontaine2012,fontaine2013}, where the $N$ input SMFs are tapered adiabatically into a FMF. This enables mode MUXes with very low insertion loss and mode-dependent loss. However, keeping the level of crosstalk between modes at a level compatible with uncoupled mode multiplexed systems appears extremely challenging: recent work shows that by using dissimilar SMFs, -6~dB of crosstalk can be achieved~\cite{leon-saval2014}. Thus, up to now, this last type of MUX has been used mainly for strongly coupled mode multiplexed transmission where mode demultiplexing is done by signal processing at receiver end. 

There also exist many implementations of mode converters and mode couplers directly in waveguides, which can be easily integrated on chip: long period fiber grating converters~\cite{giles2012}, multimode intereference couplers~\cite{leuthold1996}, asymmetric or symmetric fused fiber couplers~\cite{ding2013,li2013}, asymmetric Y-junctions~\cite{love2012,driscoll2013}, or microring resonators~\cite{luo2014}. All these techniques require very strict longitudinal phase-matching conditions, therefore limiting their bandwidth of operation; their compatibility with wavelength-division multiplexing is still under consideration.

In this article we propose and demonstrate a new type of mode multiplexer based on a succession of transverse phase profiles, which can be configured to perform any unitary spatial transform. This Multi-Plane Light Converter (MPLC) is able to achieve multiplexing to any set of spatial modes, with low intrinsic loss whatever the number of modes while achieving a very low level of crosstalk, for a broad wavelength range from 1530 to 1565~nm.

\section{Unitary spatial multiplexer}

\begin{figure}[h]
\centering
\includegraphics[scale=0.5]{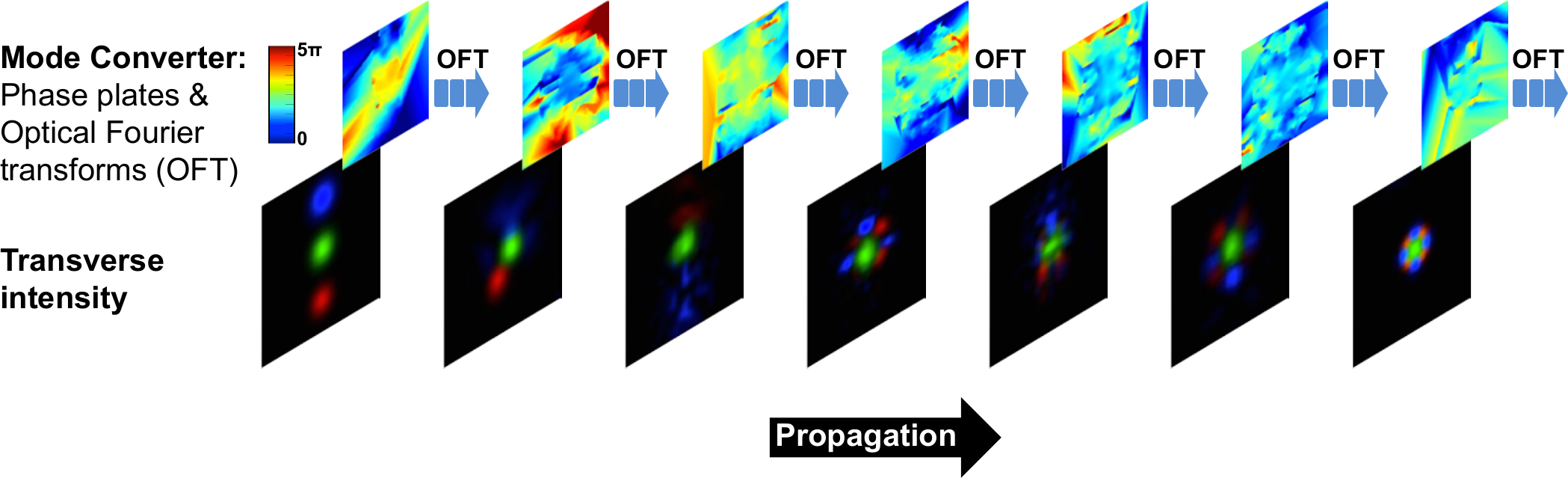}
\caption{Multi-Plane Light Converter (MPLC) implementing a spatial multiplexer, represented in transmission. Top line shows the succession of the MPLC components (transverse phase profiles and optical Fourier transforms), and the second line the evolution of the transverse intensity in the plane of the phase plate.} \label{fig:propagation}
\end{figure}

Fundamentally, spatial multiplexing is a transform that takes N separate input beams and turns them into N orthogonal modes of a few-mode fiber. Since spatial multiplexing transforms a mode basis into another mode basis, it can be considered as a spatial unitary transform~\cite{siegman1986}. We have previously shown in \cite{morizur2010} that, theoretically, for any desired spatial unitary transform between an input and output plane, there is a succession of transverse phase profiles that, separated by optical Fourier transforms, achieves the desired unitary transform. This succession of phase profiles is the Multi-Plane Light Converter. An example is presented in Fig.~\ref{fig:propagation}. Transverse phase profiles are generally built using phase plates or spatial light modulators (SLMs), used either in transmission, like pictured in Fig.~\ref{fig:propagation}, or in reflection.

The full theoretical decomposition of a unitary transform into a succession of phase profiles in general requires an unpractical number of elements, on the order of $17(N^2-N)$, where $N$ is the number of pixels chosen in the transform description~\cite{morizur2010}.  On the other hand, when we consider practical situations, with $K$ specified inputs / outputs, we have shown empirically that direct optimization yields efficient approximate results with typically $2K+1$ phase profiles~\cite{morizur2010}. Thus, for the transformation of 3 input modes into 3 outputs, 7 phase profiles are typically needed. Since this approach consists in performing a unitary transform on light with lossless elements (phase plates), the multiplexing / demultiplexing operation with a MPLC does not imply any intrinsic loss, as opposed to conventional approaches with beamsplitters. Losses in a MPLC occur due to imperfect optical elements (coating, surface roughness).

Furthermore, any set of orthogonal spatial modes can be specified as the output of a MPLC. In particular, it is possible to choose the individual eigenmodes of a FMF as the output copropagating modes. In Fig.~\ref{fig:propagation}, three of such output modes are represented (LP$_{01}$, LP$_{21\mathrm{a}}$ and LP$_{21\mathrm{b}}$). An identical MPLC, used in the reverse direction, performs the inverse unitary transform, and therefore demultiplexes the same modes.


\section{Experimental implementation and results}

\begin{figure}[h]
\centering
\includegraphics[scale=0.5]{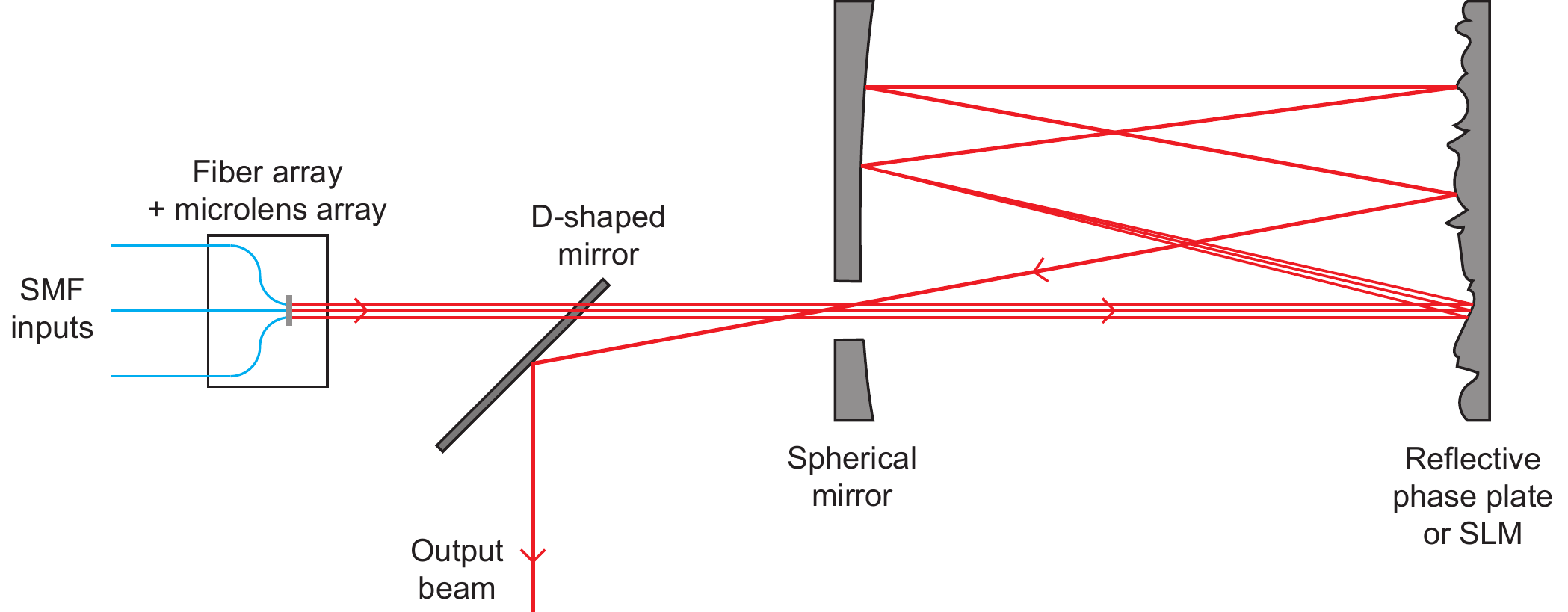}
\caption{Experimental setup of the MPLC in a multipass cavity (simplified in this representation to 3 reflections on the phase plate).}\label{fig:setup}
\end{figure}

The MPLC is experimentally implemented using a multipass cavity. The successive transverse phase profiles are all printed on a single reflective phase plate, each phase profile being located on a different spot of the plate. Optical transforms close to optical Fourier transforms (OFTs) are obtained by propagation and reflections on a spherical mirror. The cavity formed by the mirror and the phase plate allows to perform the successive phase profiles and optical transforms; a hole in the spherical mirror allows the beams to enter and exit the cavity (see Fig.~\ref{fig:setup}). 

In order to evaluate the capability of the MPLC as a spatial MUX, we use 7 reflections on the phase plate to multiplex from one to three SMF inputs into the eigenmodes LP$_{01}$, LP$_{02}$, LP$_{11\mathrm{a}}$, LP$_{11\mathrm{b}}$, LP$_{21\mathrm{a}}$ and LP$_{21\mathrm{b}}$ of a FMF~\cite{sillard2011}. 

For experimental flexibility, the phase plates are generated on a reflective liquid crystal on silicon SLM (from Boulder Nonlinear Systems); the SLM has an active surface of $512\times 512$ pixels of size $15\times 15~\mu$m, and a phase dynamics of $\sim 3\pi$ over 65536 phase levels at 1550~nm. Three single-mode fibers (SMF) are connected to a fiber array; at the output of the bundle of fibers, a microlens array is used to collimate the three beams in free space. The beams are then sent to the multipass cavity; at the output of the cavity, a D-shaped mirror is used to extract the output beam. The experimental setup is shown in Fig.~\ref{fig:setup}.

\begin{figure}[h]
\centering
\includegraphics[scale=0.5]{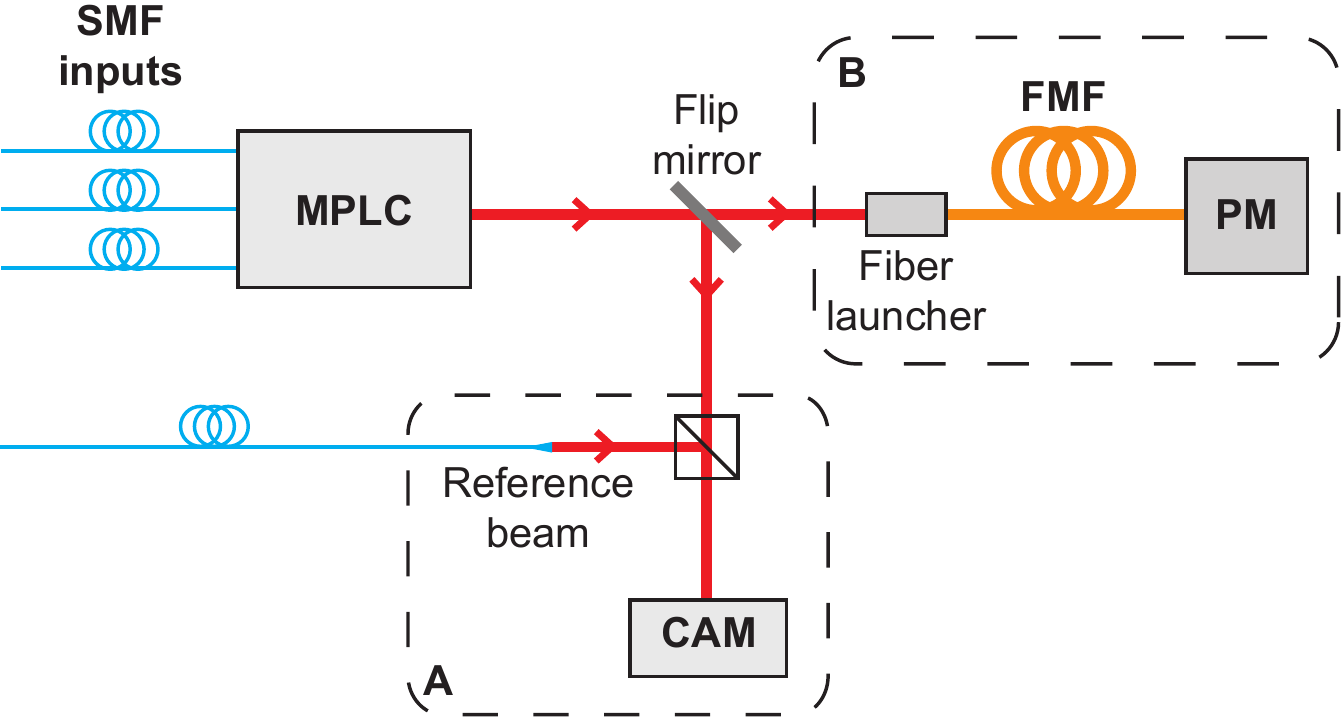}
\caption{Schematics of the two characterization setups of the produced modes. A: field measurement in free space by Fourier-transform interferometry. B: measurement of the energy coupled in the few-mode fiber. CAM: camera; PM: powermeter. The flip mirror is used to switch between the two characterization systems.}\label{fig:characterization}
\end{figure}

The modes are first characterized at the output of the multipass cavity, in free space. We measure their field by Fourier-transform interferometry~\cite{takeda1982} with a reference beam, therefore accessing both amplitude and phase of each generated mode. The characterization setup is shown in Fig.~\ref{fig:characterization}.

For each mode conversion, the overlap $\rho_{xx}$ between the (normalized) produced mode $w_p(\vec{r})$ and a theoretical mode LP$_{xx}$ of the FMF $w_{\mathrm{LP}_{xx}}(\vec{r})$ is given by:
\begin{equation}
\rho_{xx} = \left|\int w_p^*(\vec{r}) w_{\mathrm{LP}_{xx}(\vec r)} d\vec{r}\right|^2
\end{equation}
The overlap between the produced mode and the desired theoretical mode gives the fidelity of the conversion, while the overlap with the other theoretical modes of the conversion gives the crosstalk of the produced mode. The fidelity and the crosstalks of the produced output modes are shown in Table~\ref{tab:slm_before}. Typically, the setup achieves a conversion fidelity above $90\% = -0.46~\mathrm{dB}$ for one mode conversions, of $86\% = -0.64~\mathrm{dB}$ for two mode conversions, and of $84\% = -0.75~\mathrm{dB}$ for three modes conversions. We measure that all crosstalks are below $-24$~dB.

\begin{table}[h]
\centering
\caption{Performance of mode conversions using the SLM. The modes are measured by interferometry in free space after the MUX.}\label{tab:slm_before}
\begin{tabular}{|c|c|c|c|}
\hline
Conversion & Desired output modes & Conversion fidelity & Maximum crosstalk\\
type & LP$_{xx}$ &  & with theoretical modes\\
\hline
\hline

1 mode & LP$_{02}$ & $89\%$ & / \\
\hline
\hline

 & LP$_{01}$, LP$_{02}$ & $92\%$, $76\%$ & $0.09\% = -30$~dB \\
\cline{2-4}
2 modes & LP$_{11\mathrm{a}}$, LP$_{11\mathrm{b}}$ & $88\%$, $89\%$ & $0.13\% = -29$~dB \\
\cline{2-4}
 & LP$_{21\mathrm{a}}$, LP$_{21\mathrm{b}}$ & $88\%$, $85\%$ & $0.02\% = -37$~dB \\
\hline
\hline

3 modes & LP$_{01}$, LP$_{21\mathrm{a}}$, LP$_{21\mathrm{b}}$ & $93\%$, $83\%$, $83\%$ & $0.11\% = -30$~dB \\
\cline{2-4}
 & LP$_{02}$, LP$_{11\mathrm{a}}$, LP$_{11\mathrm{b}}$ & $78\%$, $80\%$, $89\%$ & $0.32\% = -25$~dB \\
\hline
\end{tabular}
\end{table}

\begin{figure}[h]
\centering
\includegraphics[scale=0.5]{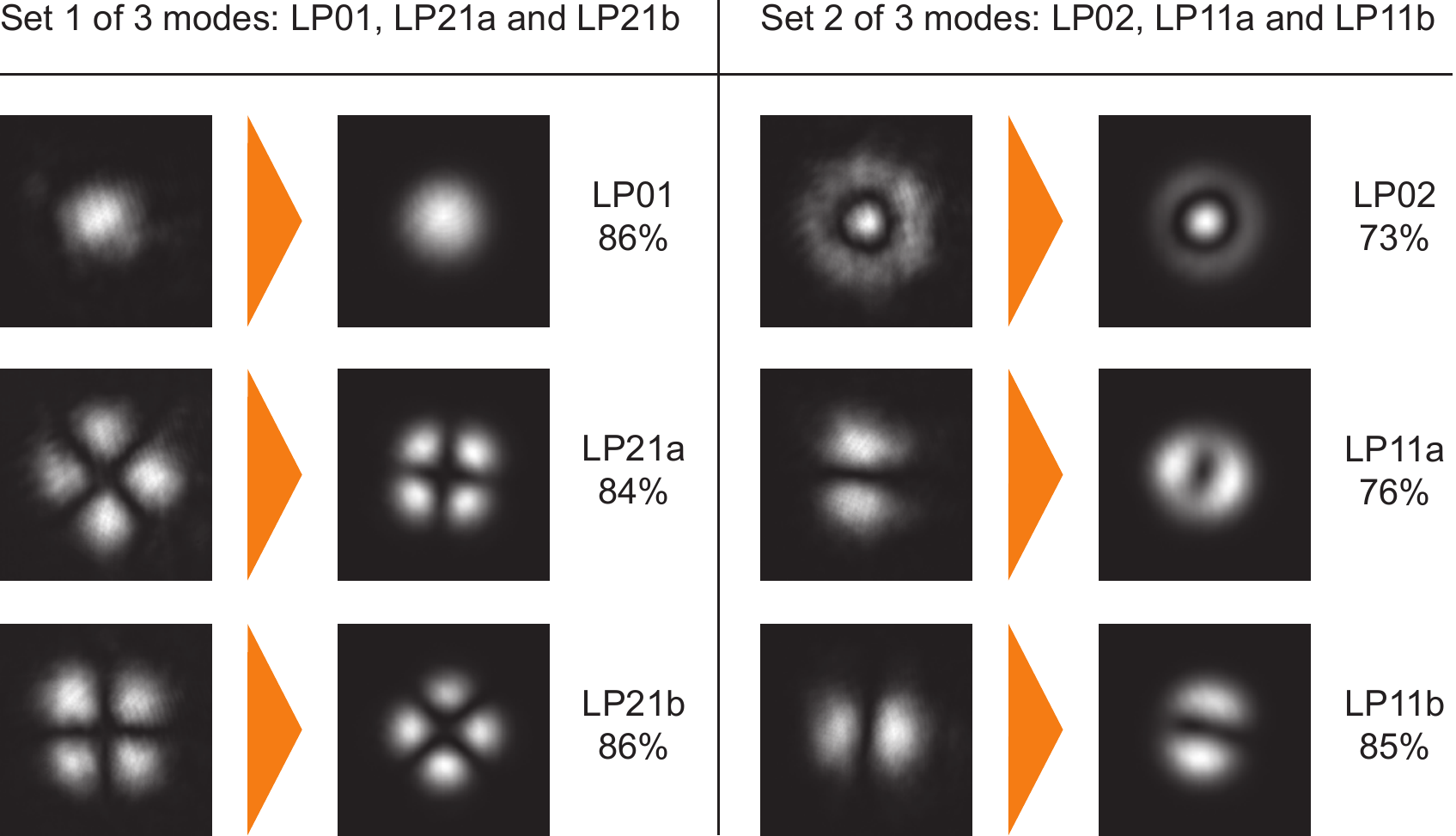}
\caption{Intensity profiles of the modes produced by the MPLC. On each column, left, before injection in the FMF; right, after propagation in 3~m of FMF. The modes are produced in a 3-mode configuration. For each mode, the fraction of energy coupled in the FMF is given.}\label{fig:injection}
\end{figure}

A second step in the characterization of the modes produced by the MPLC is the coupling of the modes in the FMF. A pair of lenses is used in a telescope configuration in order to produce a collimated output beam with a beam diameter of $12~\mu$m, corresponding to the core size of the FMF~\cite{sillard2011} (see Fig.~\ref{fig:characterization}). This characterization has been performed in a 3-mode multiplexer configuration, with the sets $\{$LP$_{01}$, LP$_{21\mathrm{a}}$, LP$_{21\mathrm{b}}\}$ and $\{$LP$_{02}$, LP$_{11\mathrm{a}}$, LP$_{11\mathrm{b}}\}$. In average, we measure that $82\% = -0.88$~dB of the energy of each produced mode is coupled into the FMF. Compensating for the Fresnel reflection losses at the input and output of the FMF, this matches the fidelity measured in Table~\ref{tab:slm_before}. In Fig.~\ref{fig:injection}, the intensity profiles of the modes produced by the MPLC, before and after the injection in the FMF, are shown.

These implementations of the MPLC, using a SLM as a reflective phase plate, show that this technology is indeed able to address the individual eigenmodes of a few-mode fiber, with high fidelity and very low crosstalk between the modes. However, the use of a SLM is not suited for a realistic implementation of a spatial MUX for telecommunications. Indeed the SLM acts as a phase plate only on one linear polarization, the orthogonal polarization being insensitive to the phase applied by the liquid crystals. It would therefore be more complex to make it compatible with current multiplexing technologies for optical fiber transmission. Moreover, the reflectivity of the SLM is of $92.5\%$, therefore it introduces a large amount of optical losses in the system. Finally, the high cost of the SLM would be an issue for the widespread adoption of this multiplexing technique.

In order to have a more suited device for mode multiplexed transmission experiments, the SLM is replaced by a reflective silica phase plate for a three mode conversion (LP$_{01}$, LP$_{21\mathrm{a}}$, LP$_{21\mathrm{b}}$). The phase plate is obtained by iterative photolithography. In total, $2^7=128$ phase levels of depth $10$~nm are obtained; the spatial resolution of the phase plate is 10~$\mu$m. The phase plate is then coated with gold in order to make it reflective at 1550~nm. The quality of the produced modes is measured with the same techniques as for the modes produced with the SLM. We measure before injection in the FMF a conversion fidelity of $84\%$ for modes LP$_{21\mathrm{a}}$ and LP$_{21\mathrm{b}}$, and of $94\%$ for mode LP$_{01}$. About $83\%$ of the energy in the LP$_{21}$ modes is coupled in the FMF, and $91\%$ of the energy of the LP$_{01}$ mode is coupled (see Table~\ref{tab:phaseplate}). The maximal crosstalk with the other theoretical modes of the conversion is below $0.59\% = -22$~dB. Moreover, we measure the power transmission of the whole multiplexer device. We observe that about $50\% = -3$~dB of the initial power in the SMF is lost during the transmission in the system. The majority of these optical losses are inside the multipass cavity, and are due to the coating on the optical elements: indeed, the gold coating on the phase plate has a reflectivity of $96\%$, and the mirrors of the cavity have a reflectivity of $99\%$; given the number of reflections inside the cavity, this leads to a total transmission of the cavity of $60\%$. In total, the addition of imperfect mode conversion and optical losses amounts to a total coupling in the FMF of $39\%=-4.1$~dB for modes LP$_{21\mathrm{a}}$ and LP$_{21\mathrm{b}}$, and of $49\%=-3.1$~dB for mode LP$_{01}$ (see Table~\ref{tab:phaseplate}).

\begin{table}[h]
\centering
\caption{Performance and losses of the MPLC for a 3 mode conversion from 3 SMFs to LP$_{01}$, LP$_{21\mathrm{a}}$ and LP$_{21\mathrm{b}}$, using a reflective phase plate. The coupling loss in the FMF takes into account the Fresnel reflection loss at the input and the output of the FMF.}\label{tab:phaseplate}
\begin{tabular}{|c|c|c|c|}
\hline
 & LP$_{01}$ & LP$_{21\mathrm{a}}$ & LP$_{21\mathrm{b}}$ \\
\hline
Optical loss before & & & \\
the multipass cavity & 0.76~dB & 0.81~dB & 0.86~dB  \\
\hline
Optical loss in &  & & \\
the multipass cavity & 1.89~dB & 2.34~dB & 2.55~dB \\
\hline
\multicolumn{4}{c}{}\\
\hline
Conversion loss &  & & \\
in free space & 0.28~dB & 0.84~dB & 0.67~dB \\
\hline
\hline
Total conversion loss & & & \\
in free space & 2.93~dB & 3.99~dB & 4.08~dB \\
\hline
\multicolumn{4}{c}{}\\
\hline
Coupling loss & & & \\
in the FMF & 0.43~dB & 0.88~dB & 0.69~dB \\
\hline
\hline
Total coupling loss &  & & \\
in the FMF & 3.08~dB & 4.03~dB & 4.10~dB\\
\hline
\end{tabular}
\end{table}

Let us stress that the intrinsic losses only amounts to a maximum of $-0.88$~dB for LP$_{21}$ modes, and to $-0.43$~dB for LP$_{01}$. The optical losses can be largely reduced using better coatings on the optical elements. Using dielectric coatings and higher reflectivity mirrors, total coupling on the order of $-2.4$~dB should be achievable. The coatings also introduce a small polarization dependent loss, on the order of $-0.5$~dB, which can be reduce by using optical coatings which have the same reflectivity for all polarizations. Apart from this effect, the MPLC is polarization independent. We have verified experimentally that it preserves the polarization of the input beams during the mode conversion.

\begin{figure}[h]
\centering
\includegraphics[scale=0.3]{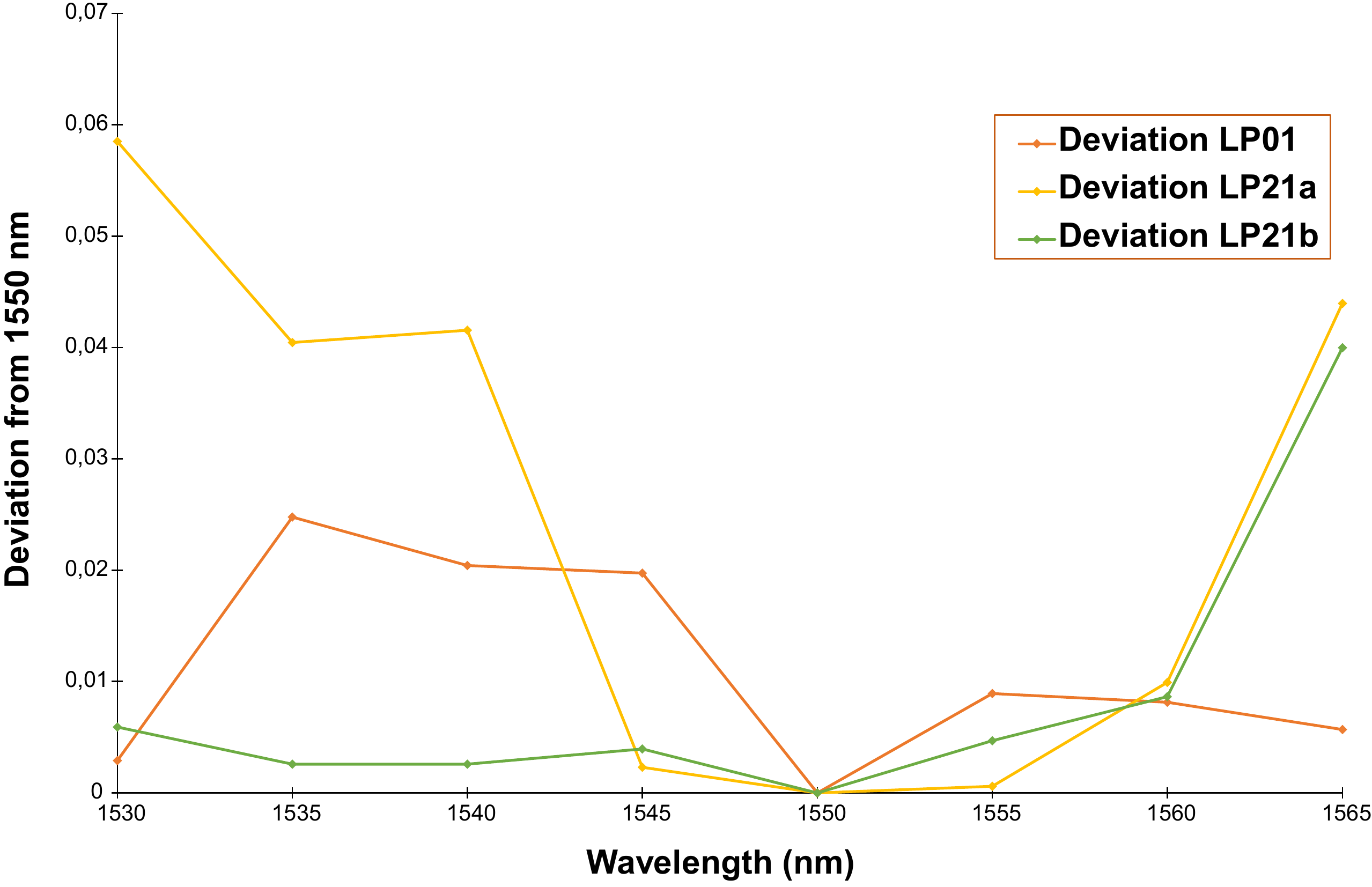}
\caption{Deviation of the coupling loss compared to the performance at 1550~nm across the C-band.}\label{fig:c-band}
\end{figure}

We have also measured the performance of the MPLC across the full C-band (1530 to 1565~nm). For 8 different wavelengths separated by 5~nm, the coupling loss in the FMF and the maximum crosstalk have been measured for the modes generated by the silica phase plate (see Fig.~\ref{fig:c-band}). We show that the deviation of insertion loss across the C-band is below $6~\%$ for all modes, and that the variation of the maximum crosstalk is of 1~dB, therefore demonstrating the full capability of the system for a large bandwidth.

Finally, we have also tested the MPLC in a DEMUX configuration. One MPLC implemented with a SLM is used as the MUX for the modes LP$_{01}$, LP$_{21\mathrm{a}}$ and LP$_{21\mathrm{b}}$, which are injected in a FMF. After a short propagation of 5~m in the FMF, the modes are demultiplexed with the MPLC implemented with the silica phase plate. The overall system MUX / FMF / DEMUX is characterized by measuring the output power couplings and crosstalks.
We measure a transmission of $-6.9$~dB for the input beam converted into LP$_{01}$, and of $-8.9$~dB for the input beams converted into LP$_{21}$ modes. The crosstalk, defined as the total power in unwanted outputs divided by the power in desired output, reaches a maximum of $-20$~dB. These figures confirm the high mode selectivity of the MPLC.

\section{Conclusion}

We have demonstrated in this article a new mode-selective spatial multiplexer / demultiplexer based on a Multi-Plane Light Converter. This device achieves one of the highest reported mode selectivity ($>23$~dB) with a coupling loss on the order of other mode selective MUXes ($<4.1$~dB) across the full C-band. This device can work as a MUX or a DEMUX with the same performance. Moreover, the intrinsic loss of the system, i.e. the imperfection of the mode conversion, is low ($<1.2$~dB), and the optical losses could be decreased with better coatings on the optical elements (for example, by using dielectric coatings instead of gold coatings). The MPLC is therefore fully compatible with a wavelength- and space-division multiplexed optical transmission line.

Finally, the technique of the MPLC for mode conversion is able to address any spatial mode profile with high fidelity, and can be applied to multiplex a higher number of spatial modes by increasing the number of reflections in the multipass cavity. With minor changes in the design of the system, this device can support up to 6 modes. Therefore, the MPLC proves to be a scalable technology for the implementation of a flexible mode selective spatial MUX / DEMUX, with low crosstalk and high efficiency.


\end{document}